\documentclass[aip,apl,reprint,showpacs,twocolumn,superscriptaddress,floatfix]{revtex4-1}

\usepackage{epsfig}
\usepackage{amsmath}
\usepackage{amssymb}
\usepackage{amsfonts}
\usepackage{mathptmx}
\usepackage{eucal}
\usepackage{bm}
\usepackage{color}
\usepackage[colorlinks,linkcolor=blue,citecolor=blue]{hyperref}
\usepackage{subfigure}
\usepackage{graphicx}
\usepackage{natbib}

\begin{document}

\title{Superconducting cascade electron refrigerator}

\author{M. Camarasa-G\'{o}mez}
\affiliation{NEST, Instituto Nanoscienze-CNR and Scuola Normale Superiore, I-56127 Pisa, Italy}
\author{A. Di Marco}
\affiliation{LPMMC-CNRS, Universit\'{e} Joseph Fourier, 25 Avenue des Martyrs, 38042 Grenoble, France}
\author{F. W. J. Hekking}
\affiliation{LPMMC-CNRS, Universit\'{e} Joseph Fourier, 25 Avenue des Martyrs, 38042 Grenoble, France}
\author{C. B. Winkelmann}
\affiliation{Univ. Grenoble Alpes, Institut N\'eel, F-38042 Grenoble, France}
\affiliation{CNRS, Institut N\'eel, F-38042 Grenoble, France}
\author{H. Courtois}
\affiliation{Univ. Grenoble Alpes, Institut N\'eel, F-38042 Grenoble, France}
\affiliation{CNRS, Institut N\'eel, F-38042 Grenoble, France}
\author{F. Giazotto}
\affiliation{NEST, Instituto Nanoscienze-CNR and Scuola Normale Superiore, I-56127 Pisa, Italy}

\date{\today}

\begin{abstract}
The design and operation of an electronic cooler based on a combination of superconducting tunnel junctions is described. The cascade extraction of hot-quasiparticles, which stems from the energy gaps of two different superconductors, allows for a normal metal to be cooled down to about 100 mK starting from a bath temperature of 0.5 K. We discuss the practical implementation, potential performance and limitations of such a device.
\end{abstract}

\maketitle

Electronic heat transport at the mesoscopic scale has been in the spotlight during the last few years.\cite{GiazottoRev} In particular, efforts have been made to develop different types of solid-state electronic refrigerators based on tunnel junctions between a normal metal and superconductors.\cite{MuhonenRPP12} Since the first observation of electronic cooling,\cite{NahumAPL94} different kinds of devices have been studied such as SINIS and S$_2$IS$_1$IS$_2$, where S$_1$ and S$_2$ are different superconductors, N is a normal metal and I stands for a tunnel barrier. Symmetric structures avoid the use of any other contact than the cooling junctions, while the cooling power, being an even function of the voltage, is doubled. In every case, electronic cooling is obtained by applying a voltage bias related to the gaps of the superconductors S$_{1,2}$. This allows the extraction of hot quasiparticles from N to S$_1$ or from S$_1$ to S$_2$ in a SINIS or a S$_2$IS$_1$IS$_2$ structure respectively, so that as a whole cooling occurs in the normal metal or the low-gap superconductor. Such devices are of wide interest for cooling microscopic\cite{ClarkAPL05} as well as macroscopic objects.\cite{LowellAPL13}

In the SINIS case, the cooling power is maximum at a temperature around $T_c/3$, where $T_c$ is the superconducting critical temperature. By exploiting aluminum (Al) as superconducting material with a critical temperature of about 1 K, this optimum occurs at a bath temperature of about 300 mK, and electronic cooling down to below 100 mK of a, for instance copper (Cu), island can be routinely achieved.\cite{MuhonenRPP12} The cooling of a superconductor by quasiparticle tunneling in a S$_2$IS$_1$IS$_2$ has also been demonstrated using aluminum-oxide-titanium junctions.\cite{ManninenAPL99} Operation over a wider temperature range calls for the use of alternative superconducting materials and/or new architectures. For instance, a SIS'IS nanorefrigerator based on vanadium (V) with a critical temperature of about $\sim 4$ K was used to efficiently cool down electrons in an Al island from 1 K to 0.4 K.\cite{QuarantaAPL}

In this Letter, we  theoretically discuss the feasibility and performance of a multistage superconducting refrigerator, hereafter called \emph{cascade} cooler. By using suitable materials and device parameters, we show that it is possible to cool down a normal metal with improved performance with respect to more conventional SINIS refrigerators.

\begin{figure}[!t]
\includegraphics[width=\columnwidth]{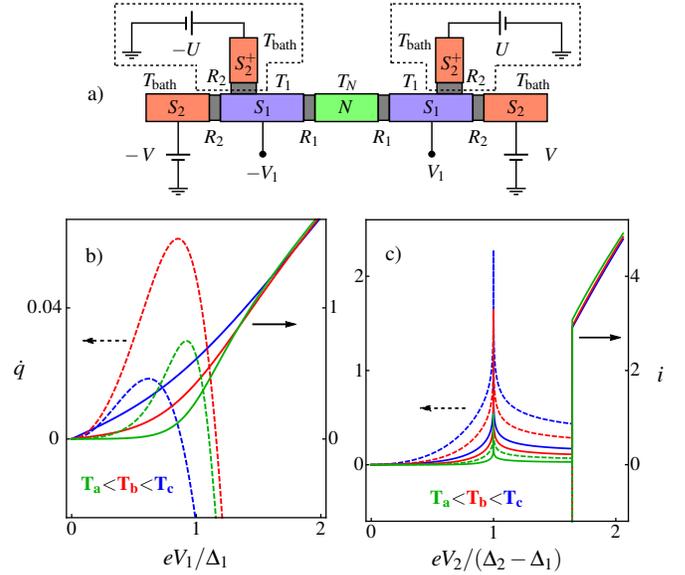}
\caption{(a) S$_2$IS$_1$INIS$_1$IS$_2$ cascade cooler geometry. The optional elements contained into the two dashed boxes enable to reach precisely the optimum bias in both the NIS$_1$ and the S$_2$IS$_1$ junctions. (b,c) Calculated dimensionless charge current $i=eR_{1,2}I_{N1,12}/\Delta_{1}$ (solid lines; right axis) and cooling power $\dot{q}=e^2R_{1,2}\dot{Q}_{N1,12}/\Delta_{1}^2$ (dashed lines; left axis) of a NIS$_1$ and a S$_2$IS$_1$ tunnel junction as a function of the dimensionless applied bias voltage $eV_1/\Delta_{1}$ and $eV_2/(\Delta_{2}-\Delta_{1})$ for different values of $T_{bath}$. In (b) we set $k_BT/\Delta_1$ = 0.129, 0.259, 0.474, respectively for $T_a$, $T_b$, $T_c$, corresponding to temperatures 0.3, 0.6, 1.1 K in the case of Al with $\Delta_1$ = 200 $\mu$eV. In (c) we use 0.345, 0.560 and 0.689, corresponding to 0.8, 1.3 and 1.6 K. The ratio $\Delta_{2}/\Delta_{1}$ of 4.105 corresponds to the V-Al combination.}
\label{Figure1}
\end{figure}

We consider an electron cooler based on tunnel junctions arranged in a symmetric configuration, \mbox{i.e.} S$_2$IS$_1$INIS$_1$IS$_2$, as displayed in Fig. \ref{Figure1}(a). The structure includes two superconductors S$_1$ and S$_2$ with respective energy gaps $\Delta_{1,2}$ so that $\Delta_1 < \Delta_2$. $R_1$ and $R_2$ denote the normal-state resistances of the individual S$_1$IN and S$_2$IS$_1$ junctions, respectively. The present structure actually consists of a SINIS micro-cooler to which one superconducting tunnel contact has been added at each end. In the following, the cascade cooler S$_2$ electrodes are voltage-biased at a voltage $\pm V$, so that the inner superconducting islands (S$_1$) reach a voltage $\pm V_1$. Here, we also assume that inelastic electron-electron interaction drives each individual part of the the system into a quasi-equilibrium regime. Therefore, the electron populations in N and S$_1$ can be respectively described by a Fermi-Dirac energy distribution function at temperatures $T_{N}$ and $T_{1}$, which can largely differ from the bath temperature $T_{bath}$. The outer superconductor S$_{2}$ is considered at thermal equilibrium with the phonon bath so that $T_{2} = T_{bath}$.

We first discuss the behavior of each individual junction in the cascade cooler. The charge current $I_{N1}$ and the heat current $\dot Q_{N1}$ flowing from N to S$_1$ through a NIS$_1$ junction under voltage bias $V_1$ are given by \cite{GiazottoRev}
\begin{eqnarray}
I_{N1} &=& \frac{1}{eR_{1}} \int_{-\infty}^{\infty} dE n_{1}(E-eV_1) [{f_N(E)-f_1(E-eV_1)}], \\
\dot Q_{N1} &=&\frac{1}{e^2R_{1}} \int_{-\infty}^{\infty} EdE n_{1}(E-eV_1) [f_N(E)-{f_1(E-eV_1)}].
\end{eqnarray}
Here, $f_{1,2,N}$ is the quasiparticle energy distribution function in S$_1$, S$_2$ or N, respectively, and $n_{1,2}$ denotes the dimensionless BCS density of states of S$_{1,2}$ smeared by the Dynes parameter\cite{DynesPRL84} $\gamma_{1,2}\Delta_{1,2}$.

In a NIS$_1$ junction, a non-zero Dynes parameter for S$_1$ induces heating in N, so that cooling vanishes at an electron temperature $T_N \simeq T_c 2.5 \gamma_1^{2/3}$. In practice, the Dynes parameter ranges from $10^{-2}\Delta_{1,2}$ to $10^{-7}\Delta_{1,2}$.\cite{PekolaPRL10} Figure \ref{Figure1}(b) shows the voltage bias dependence of the charge and heat currents in a NIS$_1$ junction. At a sub-gap bias, the heat current $\dot Q_{N1}$ is positive, meaning heat removal from N into $S_1$. At low temperature $k_BT_N < \Delta_1$, the maximum cooling power is obtained at a voltage $eV_1\simeq \Delta - 0.66k_BT_N$.\cite{MuhonenRPP12} At this optimum value, the corresponding charge current reads
\begin{equation}\label{currN1approx}
I_{N1,opt}\approx 0.48 \frac{\sqrt{k_BT_N \Delta_1}}{eR_1}.
\end{equation}
As every tunneling event removes an energy of about $k_BT$, the related heat current $\dot Q_{N1}$ is about $I_{N1,opt}k_BT_e/e$. For $eV_1>\Delta_1$, the N electrode is heated with a power $- \dot Q_{N1}$ close to $IV_1/2$. In every case, the superconductor receives a heat $- \dot Q_{1N}=IV_1+\dot Q_{N1} > 0$.

In a $\text{S}_{2}\text{I}\text{S}_{1}$ junction biased with a voltage $V_2$, the charge current $I_{12}$ and heat current $\dot {Q}_{12}$ flowing from S$_1$ to S$_2$ are given by\cite{Frank,ManninenAPL99,Giazotto2004}
\begin{eqnarray}
I_{12}&=&\frac{1}{eR_{2}} \int_{-\infty}^{\infty} dE n_{1}(E) n_{2}(E-eV_2)\nonumber\\
&\times& [f_2(E-eV_2)-f_1(E)],\\
\dot{Q}_{12}&=&\frac{1}{e^2R_{2}}\int_{-\infty}^{\infty} EdE n_{1}(E) n_{2}(E-eV_2)\nonumber\\
&\times&[f_1(E)-f_2(E-eV_2)].
\end{eqnarray}
Figure \ref{Figure1}(c) shows the voltage bias dependence of the charge and heat current in $\text{S}_{2}\text{I}\text{S}_{1}$ case. We note the sharp maximum of thermal  and charge currents occurring at a voltage bias $V_2$ equal to $(\Delta_2-\Delta_1)/e$. This peak shows up only at non-zero temperatures and corresponds to electrons occupying states above the gap in S$_1$ tunneling to empty states below the gap in S$_2$. Both the charge and the heat current at the peak are strongly affected by the temperature and the Dynes parameter. In particular, we have calculated the charge current to be
\begin{equation}\label{curr12approx}
I_{12,opt}\approx \frac{-\sqrt{\Delta_1 \Delta_2}}{eR_2} \exp{\big[-\frac{\Delta_1}{k_BT_1}\big]} \ln{\big(\sqrt{\gamma_1}+\sqrt{\gamma_2}\big)}
\end{equation}
when $\Delta_2/\Delta_1>T_\textrm{bath}/T_{1}W>1$. It is worth emphasizing that $I_{12,opt}$ depends logarithmically on $\gamma_1$ and $\gamma_2$. Compared to the NIS case, the charge current is smaller by a factor of about $\text{exp}(-\Delta_1/k_BT_1)$. The related heat current is about $I_{12,opt}\Delta_1/e$, meaning that every tunneling event removes a heat $\Delta_1$ from S$_1$.

In a normal metal, electrons exchange heat with lattice phonons with a power \cite{WellstoodPRB93} $P_{e-ph}(T_N,T_{bath}) = \Sigma \mathcal{V_N}(T_N^5 - T_{bath}^5)$, where $\mathcal{V_N}$ is the N metal volume and $\Sigma$ is the material-dependent electron-phonon coupling constant. In a superconductor, the energy gap around the Fermi level suppresses the efficiency of the electron-phonon coupling. At $T_{bath}\ll T_{1}\ll \Delta/k_B$, one obtains that the power exchanged between electrons and phonons ($P_{e-ph}^S$) is reduced by a factor of $0.98 \exp{(-\Delta/k_B T_{1})}$ with respect to that of the normal state.\cite{TimofeevPRL09}

\begin{figure}[!t]
\includegraphics[width=\columnwidth]{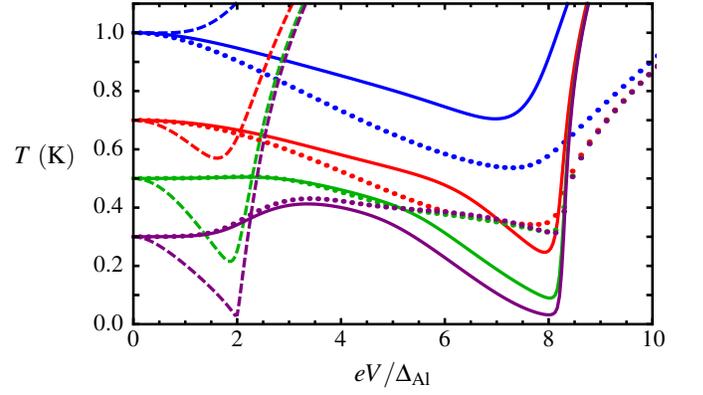}
\caption{Calculated temperature of the normal metal $T_{N}$ (solid line) and of the superconductor S$_1$ $T_{1}$ (dotted lines) for a V-Al-Cu cascade cooler and for an Al-Cu SINIS refrigerator (dashed lines) as a function of $eV/\Delta_{Al}$, at a bath temperature $T_{bath}$ = 1 K (blue curves), 0.7 K (red), 0.5 K (green), and 0.3 K (purple). The parameters are $\gamma_{1,2}=10^{-5}$, $R_1$ = 500 $\Omega$, $R_{1}/R_{2}= 100, \mathcal{V}{_1}=\mathcal{V}_{N}$ = 10$^{-2} \mu$m$^3$, $\Delta_{Al}$ = 200 $\mu$eV, $\Delta_{V}$ = 820 $\mu$eV, $\Sigma_{Al}$ = $0.2 \times 10^9\text{Wm}^{-3} \text{K}^{-5}$ and $\Sigma_{Cu}$ = $2 \times 10^9\text{Wm}^{-3} \text{K}^{-5}$.}
\label{Figure2}
\end{figure}

We now consider the whole cascade superconducting refrigerator. In the series configuration that we first consider, the charge currents flowing through all junctions are necessarily equal, so that
\begin{equation}
I_{N1} = I_{12}.
\end{equation}
The thermal balance in N reads
\begin{equation}
2\dot{Q}_{N1}+P_{e-ph} = 0,
\label{BalanceN}
\end{equation}
the factor 2 coming from the presence of two symmetric cooling NIS junctions. On the other hand, the thermal balance in each $S_1$ reads
\begin{equation}
\dot{Q}_{12}+\dot {Q_{1N}}+{P_{e-ph}^S} = 0,
\end{equation}
where we have taken into account the heat $- \dot Q_{1N}>0$ deposited by the S$_1$IN junction into the superconductor 1. The behavior of the cascade cooler is governed by the above three non-linear integral equations. It depends strongly on different parameters such as the dimensionless Dynes parameters $\gamma_{1,2}$, the N and S$_1$ volumes $\mathcal{V}_{N,1}$, the choice of the materials, the bath temperature, and the junction resistances $R_{1,2}$. As for the latter, it is crucial that the two cooling junctions NIS$_1$ and S$_1$IS$_2$ reach together their optimum cooling point at a given global bias $V$. A first naive assumption would be to assume that the currents at the optimum bias point are close to the Ohm's law value, so that the resistance balance would read $(\Delta_2-\Delta_1)/R_2=\Delta_1/R_1$. This is actually incorrect, as the current through the $S_{2}IS_{1}$ junction is far from being Ohmic and depends strongly on the Dynes parameters.

In order to be more specific, let us consider as a first combination of materials vanadium, aluminum and copper. Based on its critical temperature of about 4 K, vanadium brings a good efficiency for electronic cooling from a bath temperature around 1 K.\cite{QuarantaAPL} An aluminum island cooled in this way can reach a temperature close to the operation range of usual aluminum-based SINIS coolers. A cascade combination of V-Al$_2$O$_3$-Al and Al-Al$_2$O$_3$-Cu junctions therefore seems promising. Figure \ref{Figure2} compares the behavior of a Al-Cu SINIS refrigerator (dashed lines) to a V-Al-Cu cascade cooler (solid lines) with usual parameters values, a common tunnel resistance $R_1$ value of 500 $\Omega$ and a resistance ratio $R_1/R_2$ of 100, close to the optimum (see below). From Fig. 2, the electronic cooling of the N island (full lines) is more efficient in the cascade system, which performs well up to 0.7 K whereas the SINIS refrigerator (dashed lines) is little efficient. At a bath temperature of 1 K, the SIN stage is inefficient, while the SIS stage operates well. The capability of the cascade refrigeration scheme is illustrated by the large quasiparticle cooling obtained in S$_1$ at every bath temperature below 1 K (dotted lines).

\begin{figure}[!t]
\includegraphics[width=\columnwidth]{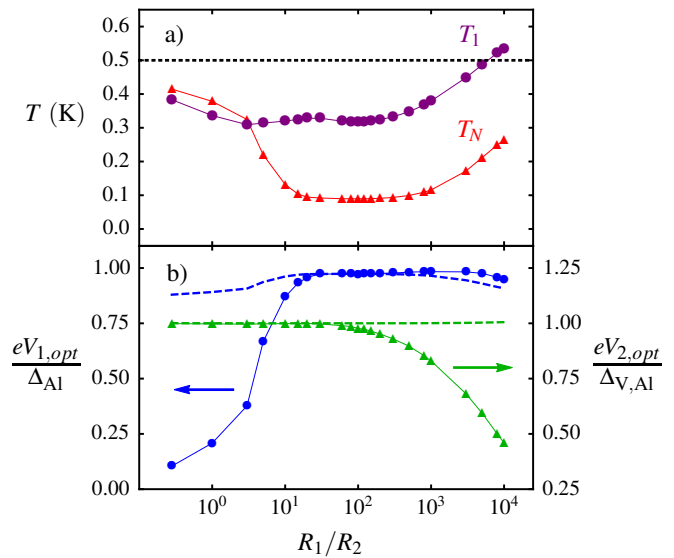}
\caption{(a) Calculated minimum temperature of the normal metal $T_{N,min}$ (red triangles) and the related temperature of low-gap superconductor $T_1$ (purple disks) of a V-Al-Cu cascade cooler at its optimum bias point as a function of the ratio $R_{1}/R_{2}$ for a bath temperature $T_{bath}$ = 0.5 K. (b) Related dimensionless voltage drops $eV_{1,opt}/\Delta_{Al}$ (blue; left axis) and $eV_{2,opt}/\Delta_{V,Al}=eV_{2,opt}/(\Delta_{V}-\Delta_{Al})$ (green; right axis) across the S$_1$IN and S$_2$IS$_1$ junctions, respectively, as a function of $R_{1}/R_{2}$. The bath temperature considered here is 0.5 K. The other parameters are identical to the ones of Fig. 2. Also shown are the predictions $eV_{1,opt}=\Delta_{1}(T_{1})-0.66 k_B T_{N}$ (dashed blue line) and $eV_{2,opt}=\Delta_{2}(T_{bath})-\Delta_{1}(T_{1})$ (dashed green line).}
\label{Fig3}
\end{figure}

Still in the case of a V-Al-Cu device, Figure \ref{Fig3} displays the minimum achieved electronic temperature in N ($T_N$) [panel (a)] and the voltage drops $V_{1,opt}$ and $V_{2,opt}$ [panel (b)] across the two S$_1$IN and S$_2$IS$_1$ junctions at the minimum temperature $T_N$ versus the junctions' resistance ratio $R_{1}/R_{2}$. A bath temperature $T_{bath}$ of 0.5 K and a fixed resistance $R_1$ of 500 $\Omega$ is considered here. At large $R_1/R_2$ value, the S$_1$IN junctions dominate and the optimum cooling is obtained at a voltage drop $V_1$ close to the expected value $(\Delta_1 - k_BT_N)/e$. At small $R_1/R_2$ value, it is the S$_2$IS$_1$ junctions that dominate, and the optimum cooling is obtained at $V_2$ close to the expectation $(\Delta_2-\Delta_1)/e$. Overall, the best performance is obtained in the region where the two kinds of junctions can operate close to the optimum. Here, the parameters are $\gamma_{1,2}=10^{-5}$ and $10^{-4}$ and $\mathcal{V}_{N}$ = 10$^{-2}$ $\mu$m$^3$. We have used the well-accepted material-specific values $\Sigma_{Al}$ = $0.2 \times 10^9\text{Wm}^{-3} \text{K}^{-5}$ and $\Sigma_{Cu}$ = $2 \times 10^9\text{Wm}^{-3} \text{K}^{-5}$. In this case, we achieve a good and somewhat constant performance for a resistance ratio between 10 and 200. This order of magnitude is consistent with the factor $\exp(\Delta_1/k_BT_1)$ between the currents $I_{N1,opt}$ and $I_{12,opt}$ at an identical junction resistance $R_{1,2}$. The relatively large span of this region stems from the existence of the singularity in the electric current as a function of the bias voltage. This rectifies any imbalance that might occur in the structure, similarly to what happens for an asymmetric pair of NIS junction in series.\cite{PekolaPhysica00} At higher bath temperature, the window for optimal resistance ratio gets narrower, and is slightly shifted towards lower values.

\begin{figure}[t]
\includegraphics[width=\columnwidth]{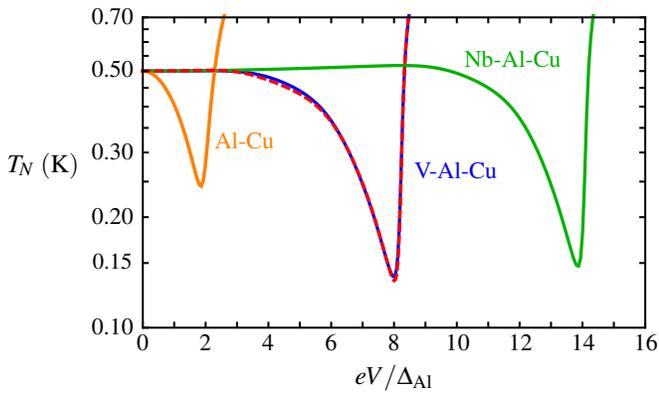}
\caption{Calculated normal metal temperature $T_{N}$ in a cascade cooler for $T_{bath}$ = 0.5 K as a function of the dimensionless bias voltage $eV/\Delta_{Al}$, in different cases: Al-Cu one-stage cooler (orange), V-Al-Cu (dotted red) with identical volumes for N and S$_1$, V-Al-Cu (blue) and Nb-Al-Cu (green) with volumes adapted to the resistances' ratio so that $\mathcal{V}_1/\mathcal{V}_N=R_1/2R_2$. The ratio $R_{1}/R_{2}$ is set at the optimal value in every case: 100 (V-Al-Cu), 30 (Nb-Al-Cu), 80 (V-Al-Cu, adapted volumes' ratio) respectively. We take $\Delta_{Nb}$ = 1407 $\mu$eV, $R_1$ = 1 k$\Omega$. The other parameters are identical to Fig. 2, including $\mathcal{V}_{N}$ = 10$^{-2} \mu$m$^3$.}
\label{Figure4}
\end{figure}

Let us now discuss practical issues in a cascade cooler's design. As stated above, the performance of the cascade cooler configuration strongly depends on the value of the ratio $R_1/R_2$. Due to the smaller value of the current $I_{12}$ through a S$_1$IS$_2$ junction compared to the current $I_{N1}$ through a S$_1$IN junction of comparable normal-state conductance, the resistance $R_2$ has to be made significantly smaller than $R_1$ in order to get an efficient cascade cooler. Optimal values of the $R_1/R_2$ ratio for bath temperatures and material configurations of experimental interest therefore lie in the range $\sim 15 - 150$, while depending strongly on subtle parameters like the Dynes parameters of $S_{1,2}$. From the fabrication point of view, it might be difficult to tune the $R_1/R_2$ ratio at its optimum with a good degree of precision. This leads to the practical necessity of tuning the voltage $V_1$ independently from the the main bias voltage $V$. One possible solution to this problem is to tunnel-couple to each S$_1$ electrode an additional superconductor S$_2'$, as shown in Fig. \ref{Figure1}(a). Biasing with a second positive (negative) voltage $U$ these two tuning junctions would enable to add (subtract) some current in the S$_1$IS$_2$ junctions compared to the S$_1$IN ones. The S$_{1}$INIS$_{1}$ current can then be tuned from zero to the double of its value at zero bias $U$. The latter limitation comes from the fact that the voltage $U$ needs to be always sub-gap in order to prevent any extra heating of the S$_1$ electrode.

For practical sample fabrication issues, one would preferably use the same tunnel barrier characteristics (in particular transparency) for the two tunnel barriers between S$_1$ on one side, and N or S$_2$ on the other side. Sticking to a particular value of the tunnel resistance ratio, and using similar thicknesses for N and S$_1$, thus leads to a volume ratio $\mathcal{V}_1/\mathcal{V}_N$ between the superconductor S$_1$ and the normal metal N approximately equal to half the inverse of the resistance ratio $R_1/R_2$. Furthermore, the values of the two superconductors' gaps can also be varied, for instance replacing vanadium with niobium (Nb). Figure \ref{Figure4} shows the results for the electron temperature $T_N$ obtained with the two materials choices V-Al-Cu, Nb-Al-Cu, at $T_{bath}$ = 0.5 K, relating or not the volumes' ratio to the resistances' ratio. The optimum resistance ratios were adjusted in every case, to respectively 30 for Nb-Al-Cu, 80 for V-Al-Cu when the volume ratio is adapted to the resistance ratio, 100 for V-Al-Cu with identical volumes $\mathcal{V}_1$ and $\mathcal{V}_N$. Imposing a larger volume $\mathcal{V}_1$ affects only slightly the performance of the whole device, with a minimum electronic temperature rising from 134 to 138 mK for the V-Al-Cu material combination. This value increases to 147 mK when V is replaced by Nb. A larger gap value does not necessarily provide an improved cooling, because it also reduces the available heat current in the S$_1$IS$_2$ junction.

Another crucial issue for the present cascade electronic cooler resides in a proper quasiparticle thermalization in the intermediate superconductor S$_1$. It is well known that superconducting-based electronic refrigerators generally suffer from poor evacuation of highly-energetic quasiparticles in the superconducting electrodes.\cite{RajauriaPRB09} To this end, quasiparticle traps of various kinds have been envisaged in order to allow their evacuation into nearby-connected normal metal layers.\cite{ONeilPRB12,NguyenNJP13} In the present design, the outer superconductor S$_2$ actually plays this role, with an increased efficiency thanks to its density of states singularity at the gap edge. An incomplete quasiparticle energy relaxation in the superconductor S$_1$ should actually not hinder the cooling in the low-gap superconductor S$_1$ compared to the present quasi-equilibrium calculations. The cascade cooler appears as rather immune against poor electronic equilibration in S$_1$. Finally, the outer superconducting electrodes S$_2$ can be efficiently thermalized through quasiparticles traps, just as it is done in the case of conventional superconducting refrigerators.\cite{GiazottoRev}

In conclusion, we have discussed a novel kind of electronic cooler based on hybrid superconducting tunnel junctions. A cascade geometry allows to cool a first superconducting stage, which is used as a local thermal bath in a second stage. The correct operation of the device strongly depends on the matching between the resistances of the the two kinds of tunnel junctions. The resulting constraint can be easily implemented in a practical device, using of a set of two additional tunnel junctions. Decoupling of local phonon population from the thermal bath \cite{PascalPRB13} in a suspended metal geometry \cite{NguyenAPL12} would improve performances compared to the situation considered here.

We acknowledge the Marie Curie Initial Training Action Q-NET no. 264034 and the EU Capacities MICROKELVIN project no. 228464 for partial financial support. We thank L. M. A. Pascal for her contribution at the very beginning of this project, A. Ronzani for help in numerics, and J. P. Pekola for fruitful discussions.

\end{document}